\documentclass[english,aps,reprint,superscriptaddress]{revtex4-2}
\usepackage[T1]{fontenc}
\usepackage[latin9]{inputenc}
\usepackage{babel}
\usepackage{amsmath}
\usepackage{graphicx}
\usepackage[unicode=true,pdfusetitle,
 bookmarks=true,bookmarksnumbered=false,bookmarksopen=false,
 breaklinks=false,pdfborder={0 0 1},backref=false,colorlinks=false]
 {hyperref}

\makeatletter
\usepackage{babel}
\usepackage{lipsum}

\usepackage{xcolor} 

\makeatother

\begin{document}
\title{Extreme events prediction from nonlocal partial information in a spatiotemporally
chaotic microcavity laser}
\author{V. A. Pammi}
\affiliation{Université Paris-Saclay, CNRS, Centre de Nanosciences et de Nanotechnologies,
91120 Palaiseau, France}
\author{M. G. Clerc}
\affiliation{Departamento de Física and Millenium Institute for Research in Optics,
Facultad de Ciencias Físicas y Matemáticas, Universidad de Chile,
Casilla 487-3, Santiago, Chile}
\author{S. Coulibaly}
\affiliation{Université de Lille, CNRS, UMR8523 - PhLAM - Physique des Lasers Atomes
et Molécules, F-59000 Lille, France}
\author{S. Barbay}
\affiliation{Université Paris-Saclay, CNRS, Centre de Nanosciences et de Nanotechnologies,
91120 Palaiseau, France}
\begin{abstract}
The forecasting of high-dimensional, spatiotemporal nonlinear systems
has made tremendous progress with the advent of model-free machine
learning techniques. However, in real systems it is not always possible
to have all the information needed; only partial information is available
for learning and forecasting. This can be due to insufficient temporal
or spatial samplings, to inaccessible variables or to noisy training
data. Here, we show that it is nevertheless possible to forecast extreme
events occurrence in incomplete experimental recordings from a spatiotemporally
chaotic microcavity laser using reservoir computing. Selecting regions
of maximum transfer entropy, we show that it is possible to get higher
forecasting accuracy using nonlocal data vs local data thus allowing
greater warning times, at least twice the time horizon predicted from
the nonlinear local Lyapunov exponent.
\end{abstract}
\maketitle
\paragraph*{}
The prediction of extreme events (EEs) occurrence, while having potentially
a large impact in many fields of science and everyday life, remains
a challenge especially in large and complex spatiotemporal systems
\citep{AkhmedievPLA11,LatifahNPG12,AlamGRL14,BirkholzPRL15,ErkintaloNP15,BayindirPLA16,GuthE19,JiangPRR22,VlachasNMI22}.
EEs, which are rare and intense amplitude phenomena --
as compared to the long-time average of an observable in a given system
\citep{Nicolis12} -- have been found in many types of systems \citep{OnoratoPR13}, either
natural or in laboratory experiments. In the latter case,
optical systems have played a great role because of the analogy between
oceanic rogue waves and optical pulses propagation in nonlinear optical
fibers \citep{SolliN07,KiblerNP10}, allowing to generate and
study these EEs with a large statistics and in
a controlled environment. EEs have also
been found in nonlinear optical dissipative systems displaying chaos
\citep{BonattoPRL11,KovalskyOL11,LecaplainPRL12,BoscoOL13,BonazzolaPRA15}
or spatiotemporal chaos \citep{MarsalOL14,SelmiPRL16,ClercOL16}.
Likewise, model-free prediction of low \citep{FarmerPRL87,AbarbanelRMP93,MaassNC02,AmilChaos19}
and high \citep{PathakPRL18,VlachasPRSA18,NakaiPRE18} dimensional
chaotic time-series have been made possible thanks to the advent of
machine learning techniques. However they usually require the precise
knowledge of the whole spatiotemporal history of a dynamical field,
which is often impossible in real situations where only a part of the dynamics is observable while some dynamical
variables remain hidden and cannot be recorded. 
When a dynamical variable is observed and used to predict the outcome of another variable, 
the concept of cross-prediction has been introduced and tested \citep{ZimmermannC18,CunilleraC19}.
The application of model-free techniques is more challenging when
dealing with experimental and natural data \citep{HamN19}, where
the resolution of the measurements in time and space is limited. Recent
results have been obtained in this area for the prediction of rogue
solitons in supercontinuum generation in an optical fiber \citep{SalmelaSR20,CoulibalyCS&F22}
and the space-time localization of of extreme wind speeds in the north
Atlantic ocean \citep{JiangPRR22}. {\color{black} In Ref. \citep{CoulibalyCS&F22},
the system is purely temporal and a spatiotemporal map is obtained
by a pseudo-space reconstruction. In Ref. \citep{JiangPRR22}, the
full spatiotemporal field is recorded and used for the forecast, thanks
to the relatively slow time scale of the system's evolution.}

In this work, we utilize a model-free reservoir computing approach
for the prediction of EEs occurrence with experimental
data from a spatiotemporal chaotic broad area laser \citep{SelmiPRL16},
where only partial information of the past spatiotemporal field is
known. The only accessible observable is the laser intensity (not
the laser material dynamics), and the dynamics can only be known accurately
and simultaneously at two given locations in space. This simulates
the common situation in practice where the spatiotemporal field is
only scarcely sampled in space. We identify the spatial locations
of {\color{black}potential} precursors using an information theoretic
measure, namely transfer entropy \citep{SchreiberPRL00}. {\color{black}
At contrast with Ref. \citep{CoulibalyCS&F22}, the precursors cannot
be identified reliably but are mostly hidden in the system's dynamical
fluctuations and in the detection noise.} A classification task is
performed using reservoir computing to identify EEs in
advance, using local and nonlocal information. We compare the prediction
results and identify regimes where the nonlocal, cross prediction
task yields better prediction accuracy than the local task.

\paragraph*{}
\begin{figure}
\includegraphics[width=1\columnwidth]{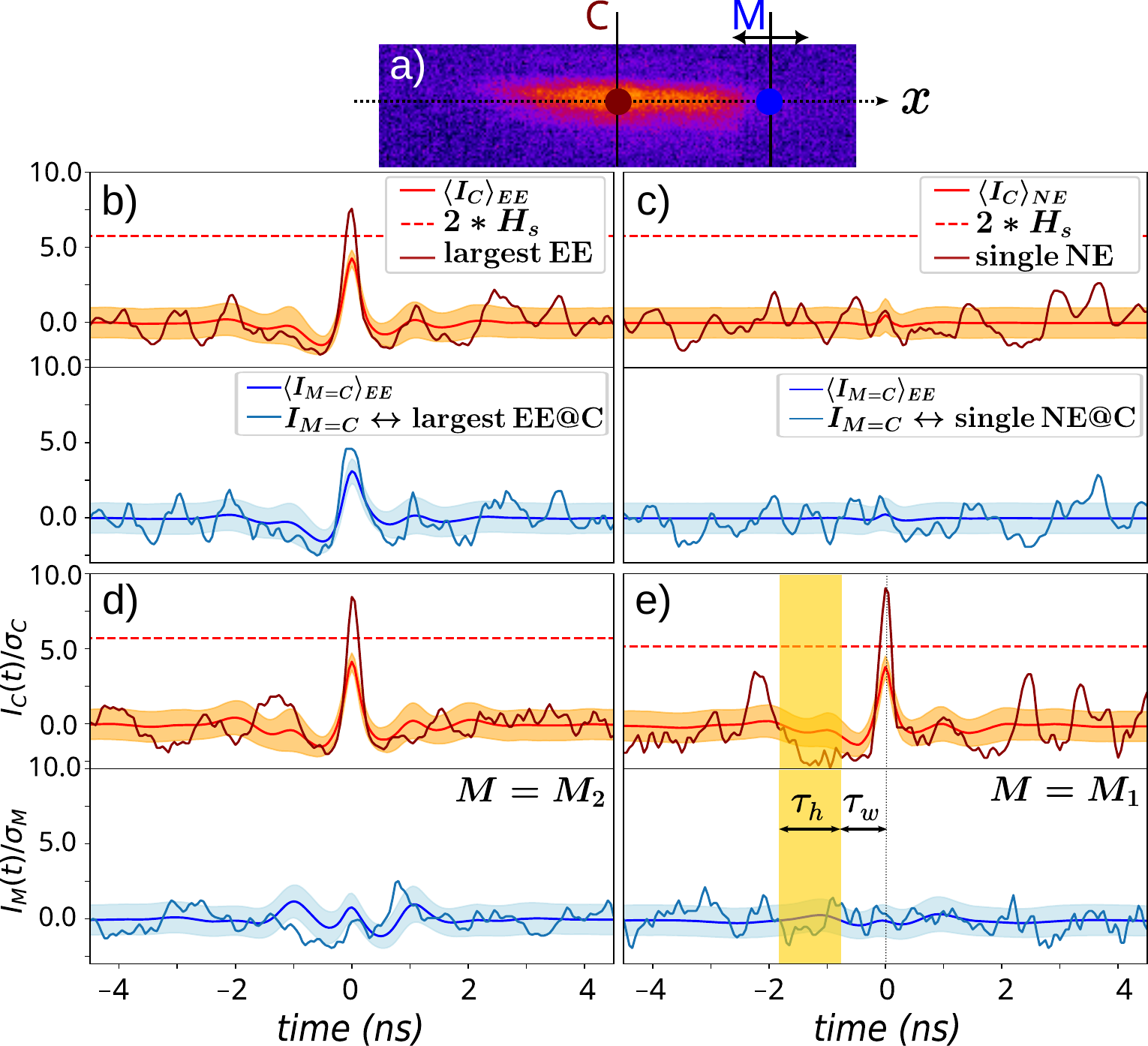}
\caption{\label{fig:correlation}a) Microlaser near-field image above lasing
threshold. b-e) simultaneous recording of scaled
intensities at C (fixed photodetector, upper) and M (mobile photodetector,
lower) for specific events at C (placed at $t=0$):
 an EE is present (b,d,e) or not present (c) at C. The photodetector is located at $x_{M}=0\ \mu$m
in b), $x_{M_{2}}=+12.3\ \mu$m in d) and $x_{M_{1}}=-13.8\ \mu$m
in e). Red and black lines: average timetraces over the ensemble of {\color{black} EE or
NE} and associated standard deviations (orange and light blue shaded
areas). Dark red: largest  EE's timetrace, in b,d,e, (resp. single random NE in
c) recorded at C  with the simultaneous timetrace recorded
at M (light blue). Red dashed line: EE threshold $2\times H_{s}$ (to be compared to the pulse height
$H$ from the trough to peak, not to the peak value). Yellow shaded
area in e): information used for the prediction of an event
at $t=0$, with history time $\tau_{h}$ and warning time
$\tau_{w}$.}
\end{figure}
We investigate a quasi-1D broad-area microcavity
laser with integrated saturable absorber which has been shown, both experimentally and numerically.
to display spatiotemporal chaos and and EEs \citep{SelmiPRL16,CoulibalyPRA17}.
The observed spatiotemporal chaos results from
 a chaotic alternation of amplitude and phase turbulence phenomena
\citep{BarbayE18}.
The microcavity laser pumped area is delimited by a clear aperture
of 10$\times$80 $\mu$m$^{2}$ and emits at $\lambda_{c}\simeq980$ nm. 
Transverse spatial coupling in the microresonator is obtained
through light diffraction with a diffraction length
 $w_{d}\simeq7.4\mu\mathrm{m}$ \citep{SelmiPRL16}.
The detailed optical setup is described in \citep{SelmiPRL16}
and recalled in the Suppl. Mat. (SM) for completeness. 
This system has the advantage of having fast timescales,
on the order of hundreds of picoseconds, thus facilitating the sampling
of a large number of low probability events in a single experimental
run. EEs are qualified using the standard definitions used
in hydrodynamics where these phenomena are coined \textquotedbl rogue
waves\textquotedbl{} \citep{OnoratoPR13}. 

The dynamics recorded at the center of the laser (see Fig.\nobreakdash-\ref{fig:correlation}a)
displays large amplitude fluctuations (Figs.\nobreakdash-\ref{fig:correlation}b,
\ref{fig:correlation}d and \ref{fig:correlation}e). These fluctuations
of height $H$ (defined as the maximum between the amplitudes measured
at the left and right sides of the pulse) can be classified into two
classes: extreme events (EE) or non-extreme events (NE). The classification
criteria for EEs is $H\ge2H_{s}$, where $H_{s}$, the
significant height, is simply the average of the height of the events
in the highest tercile. For technical reasons, it is not possible
to access the evolution of the whole section of the laser with the
required detection bandwidth. Only a partial information is available,
namely we detect the simultaneous evolution in two different points,
one fixed located at the center of the laser $I_{C}(t)=I(x_{C},t)$,
and one mobile across the transverse section $I_{M}(t)=I(x_{M},t)$.
In Fig.\nobreakdash-\ref{fig:correlation}b) the intensity of EEs
simultaneously measured by the two photodetectors at the same
location displays correlated time traces. The average time trace of
EEs shows some oscillations around the peak value at $time=0$, that
quickly dampens away from it, evidencing a typical temporal pattern
for EEs. In Fig.\nobreakdash-\ref{fig:correlation}c) by comparison,
NE are completely uncorrelated which results in a very flat average
time trace. Away from the correlation width of an EE, at $M_{1}$,
an EE recorded in C is accompanied by no clear sign in the time trace
at $M_{2}$ which displays a dynamics very similar to the one recorded
for a NE in Fig.\nobreakdash-\ref{fig:correlation}c). 
By contrast, the average signal recorded at $M=M_{2}$ (Fig.\nobreakdash-\ref{fig:correlation}d)
shows a small fluctuation for $-1.5\leq t\leq-0.5$ ns which may point
to the presence of a precursor. However, the precursor identification
is rendered difficult since the signal fluctuations are large and
on the same order of magnitude as the signal itself, as can be seen
on the non-averaged timetrace. The {\color{black}identification of
potential precursors} can be made easier using the tool of transfer
entropy described below.

\begin{figure}
\includegraphics[width=1\columnwidth]{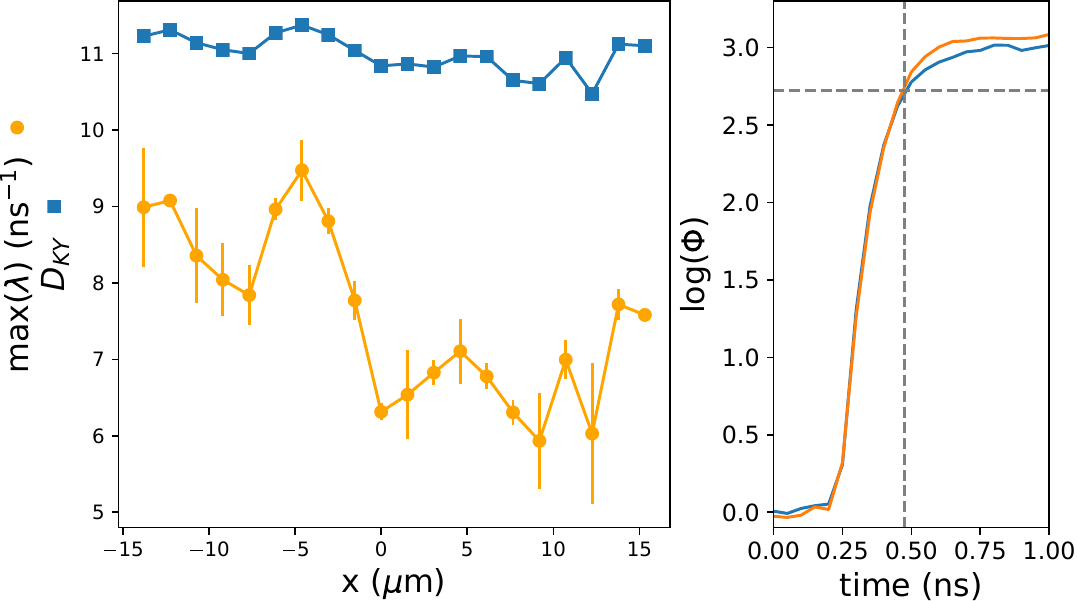}\caption{\label{fig:lyapunov}a) Largest Lyapunov exponent $\lambda_{M}=\max(\lambda)$
and Kaplan-Yorke dimension $D_{KY}$ of the time traces versus recording
position. b) Initial error growth rate for the laser intensity
versus time computed at C for both photodetector recordings. The extracted
prediction time horizon (see text) is 0.47 ns.}
\end{figure}

\paragraph*{}
The dynamical complexity of the dataset can be estimated from the
Lyapunov spectra computed for the individual local recordings $I_{M}(t)$. From
these, the largest Lyapunov exponent $\lambda_{M}$ can be extracted,
giving access to a global, mean maximum prediction time $\tau_{p}\simeq\frac{1}{\lambda_{M}}\ln(\frac{\Delta}{\delta_{0}})$
\citep{Kantz03,DingPLA07}, with $\delta_{0}$ the initial perturbation
and $\Delta$ the resolution of the measurement. The mean Kaplan-Yorke
dimension $D_{KY}$ and fractal dimension $D_{f}$ of the attractor 
are respectively $\langle D_{KY}\rangle\simeq11$ and $\langle D_{f}\rangle\simeq7.1$ (Fig.\nobreakdash-\ref{fig:lyapunov}a), which are consistent with a high-dimensional chaos. 
A more precise estimate of the
prediction time horizon is given using the rate
of growth of initial error rate $\Phi$ computed in Fig.\nobreakdash-\ref{fig:lyapunov}, which can be extracted from the nonlinear local Lyapunov exponent \citep{DingPLA07} (see SM).
The prediction horizon time can be defined as the time at which
$\log(\Phi)$ reaches 90\% of its saturation value and is of the order of $0.47$ ns here. 
Using side results (see SM) we can also estimate $\delta_{0}\simeq3\times10^{-4}$
and assuming $\Delta\simeq5\times10^{-3}$, which corresponds to a SNR of 1, and extracting the mean
maximum Lyapunov exponent from Fig.\nobreakdash-\ref{fig:lyapunov}
$\langle\lambda_{M}\rangle=7.6$ ns$^{-1}$, we get a mean maximum
prediction time $\tau_{p}=0.37$ ns. This time is slightly smaller
than the time obtained from $\log(\Phi)$,
as expected, but is still larger than the Lyapunov time usually considered
as a time horizon indicator $\tau_{L}=\langle\lambda_{M}\rangle^{-1}=0.13$
ns. Recent results showed that machine learning aided model-free predictions
of high-dimensional chaotic systems was possible up to about 6 Lyapunov
times \citep{PathakPRL18,ZimmermannC18,JiangPRR19,FanPRR20,VlachasNN20}.
\paragraph*{}
Early warning signs of EE, also called precursors, have been considered
in many previous works (e.g. in \citep{AkhmedievPLA11,BayindirPLA16,CousinsPRE19})
including in low dimensional optical systems \citep{ZamoraMuntPRA13,BonattoPRE17,AlvarezEPJST17,BonazzolaPRE18}.
{\color{black} To identify potential regions of precurors, }
we consider transfer entropy \citep{SchreiberPRL00}, similarly
as in \citep{CoulibalyCS&F22}, which 
measures the information transfer between two signals. It is more robust that e.g. a simple cross-correlation 
since it uses conditional probabilities
instead of correlations. 

\begin{figure}
\includegraphics[width=1\columnwidth]{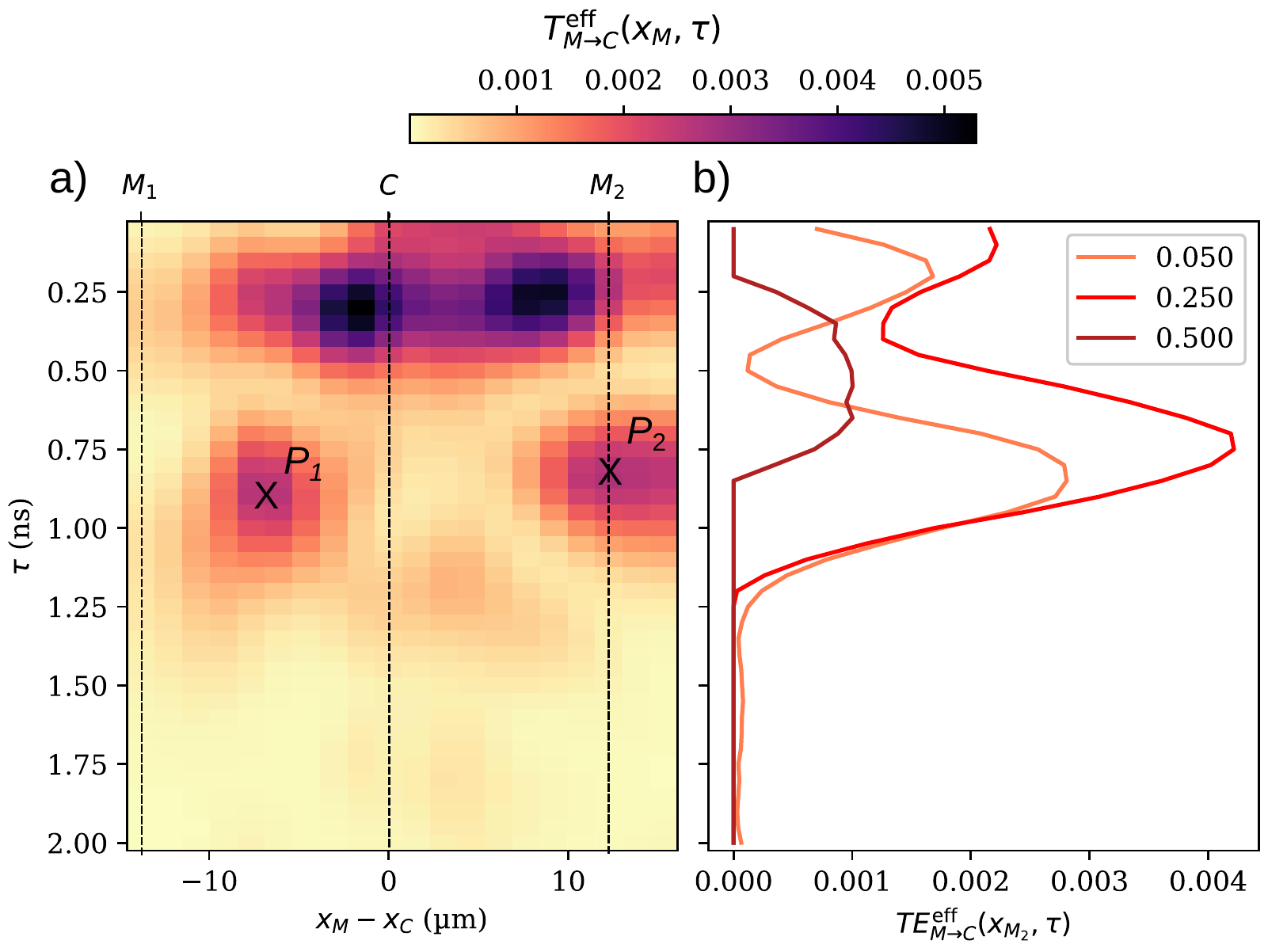}\caption{\label{fig:Transfer-Entropy}Effective transfer entropy $T_{M\to C}^{\mathrm{eff}}$
as a function of $x_{M}$ and delay $\tau$ for $\tau_{h}=1\Delta t=0.050$
ns and b) at $M_{2}$ for $\tau_{h}=0.050,\ 0.250,\ 0.500$
ns. $P_{1}$ and $P_{2}$ locate the main precursor regions
in the spatiotemporal diagram. $T_{M\to C}^{\mathrm{eff}}$
is smoothed by a small Gaussian kernel (see SM for the original data).} 
\end{figure}

We introduce 
the two-dimensional
effective transfer entropy $T_{M\to C}^{\mathrm{eff}}(x_{M},\tau)$ (see SM)
which measures the information gained at point C (in bits) from the
knowledge of a history of duration $\tau_{h}$ in the past at M (see Fig.~\ref{fig:correlation}e), with $\tau$
parametrizing the time delay in the past. 
It is obtained by subtracting to the transfer entropy $T_{M\to C}$ the
transfer entropy for surrogate data in M, allowing comparison between transfer entropies computed using different
$\tau_{h}$.
$T_{M\to C}^{\mathrm{eff}}(x_{M},\tau)$ is calculated and plotted in Fig.\nobreakdash-\ref{fig:Transfer-Entropy}a
for a history of size $\tau_{h}=0.050$ ns.  It displays three regions
of interest. A large central lobe centered around $x_{M}\simeq x_{C}$
which corresponds to causal information in the immediate spatiotemporal
surrounding of the EE, and two disconnected regions almost
symmetric about the temporal axis which we identify {\color{black}as
the location of potential precursors} (around $P_{1}$ and $P_{2}$). 
It is clearly seen that EEs extend over a finite length of $10\mu m$
width, as already noted in Ref. \citep{SelmiPRL16}. At lags around
$\tau=-5\Delta t=-0.25$ ns, there is a net transfer of information
to the center of the laser at $\tau=0$. This corresponds
to the immediate warning signal of the EE formation. More importantly,
there are disconnected regions around $P_{1}$ and $P_{2}$ at delay
times $\tau\simeq0.9$ ns where there is a net positive transfer of
information, outside of the initial correlation length of the system.
In the following, we are going to use this knowledge for a model-free
prediction of the occurrence of EE given the past dynamical information.
\paragraph*{}
A dataset is built after identifying events times $t_{E}$ of 
 intensity maxima at $C$ and recording the signal at $M$ for a 
 duration $\tau_{h}$ corresponding to $m$ samples, i.e. from $[t_{E}-\tau_w-\tau_{h},t_{E}-\tau_w]$, $\tau_w$ being
the warning time (see Fig.~\ref{fig:correlation}e).
Events at C are labelled as EE or NE. Since EEs are rare by
definition, a balanced dataset is built by retaining all the $N$
EEs and choosing an equal number of NEs
at random. This allows us to use a standard metric for the loss function
\citep{GuthE19}. The dataset therefore consists of $2\times N$ time-traces
associated to labels which identify their categories, $70\%$ of which is used as training
data and $30\%$ as testing data. The prediction
task is carried out using reservoir computing (RC). RC has been used
for prediction on various low- and high-dimensional dynamical
systems \citep{MaassNC02,JaegerS04,LukoseviciusCSR09,PathakPRL18,BianchITNNLS21}.
It is particularly interesting as reservoirs are themselves dynamical
systems making them ideal candidates
to map other dynamical systems. While we have tested other machine learning algorithms
(K-nearest neighbours, long-short term memory, logistic regression), none of
them did show a significant superiority and RC happened to be the
one with the most overall best performance \citep{PammiPhD21}. The
reservoir generation and update follow standard procedures detailed in the SM.
It comprises $N=50$ nodes each with a hyperbolic tangent
activation function and is initialized by a null state. Its parameters
were optimized thanks to a hyper-optimization routine. At the end
of the input sequence, the state of the reservoir nodes is stored
forming an output vector of length $N$. Thus, an input time-series
of $m$ samples is converted into a vector of at most $N$ values,
which is a representation of the input data. A logistic classifier
assigning a class EE or NE is then trained on all the training 
times sequences.
\begin{figure}
\includegraphics[width=1\columnwidth]{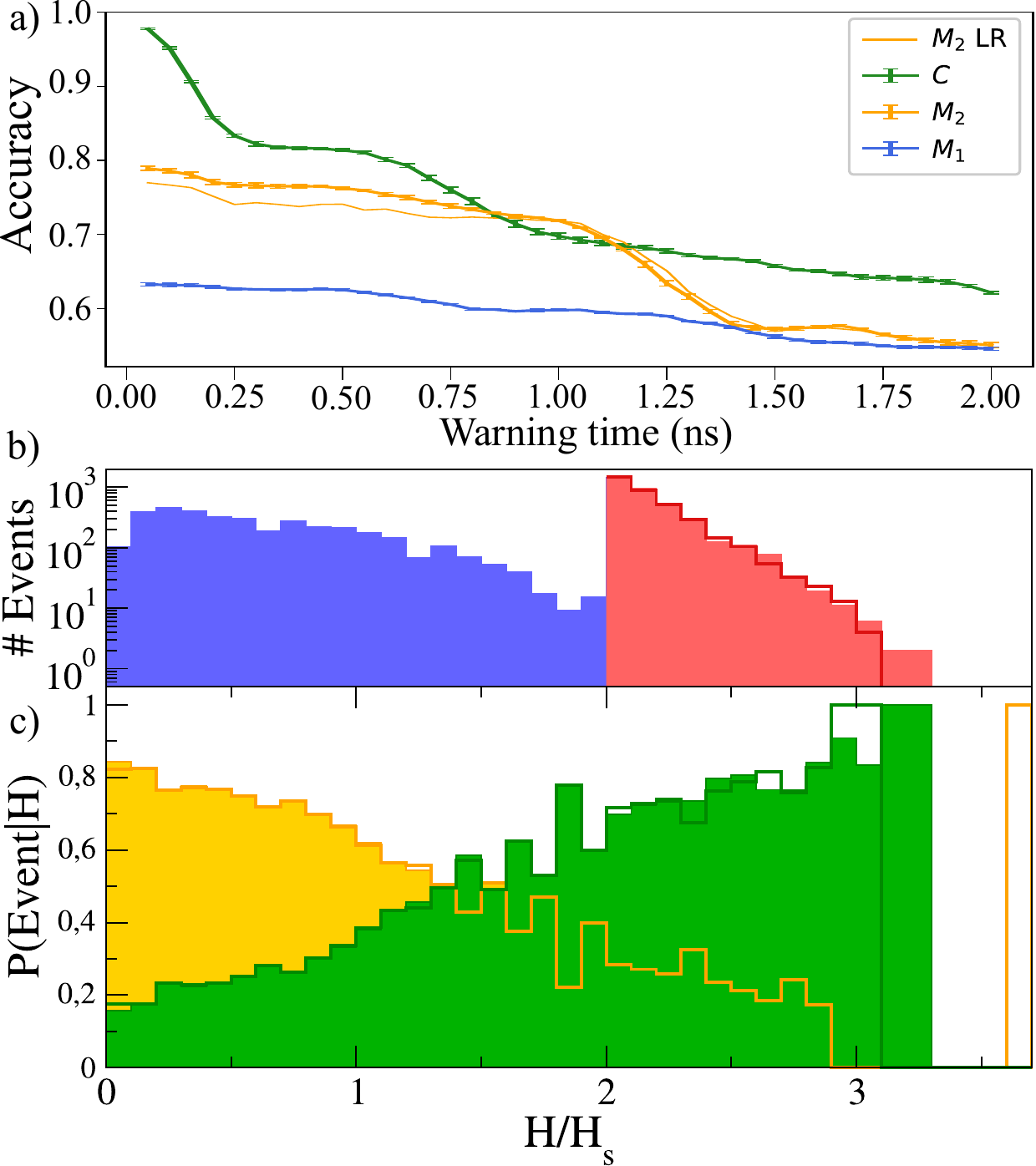}\caption{a) Mean EE forecasting accuracy at $C$ with
$\tau_h=1.75$ ns versus the warning time, using either local (at $C$) or nonlocal (at $M_{1}$ or $M_{2}$) data as input.
The mean and standard deviation are computed with 10 different realizations
of the reservoir. Thin line: nonlocal forecasting at $M_{2}$ using logistic regression alone. b) Histogram of the testing
dataset heights distribution (blue, NE and red, EE events). c) classification
probability as an EE (green) or a NE (complementary, orange) of an
event of actual height $H$ occurring $\tau_{w}=1$ ns in the future
at C from the knowledge of a history of $\tau_{h}=1.75$ ns duration
of the non-local data at $M_{2}$. Two different realizations of the
testing dataset are illustrated (plain and empty histograms). \label{fig:Training-set-PDF-1}}
\end{figure}
The forecasting accuracy of an EE at $C$ for a history $\tau_h=1.75$ ns versus the warning time $\tau_w$ is shown on Fig.\nobreakdash-\ref{fig:Training-set-PDF-1}a.
It is displayed for different training data: using the local information
at $C$, or the nonlocal data at $M_{1}$ or $M_{2}$. Note that the
accuracy does not depend significantly on the history length after
a certain length is reached (see SM). In the first case, an accuracy close to 1 is obtained for small
warning times, since this forecasting task is linear and
simple. As $\tau_w$ increases, the forecasting accuracy also
decreases almost monotonically towards 0.5, i.e. to the
absence of forecasting power. The same behavior occurs using nonlocal
training data, though with important differences between $M_{1}$
and $M_{2}$. At $M_{1}$, the forecasting accuracy is always low
since there is almost no information present at this location, as
can be checked in Fig.\nobreakdash-\ref{fig:Transfer-Entropy}. The
forecasting accuracy using nonlocal data at $M_{2}$, on the contrary,
is close to $0.8$ for small warning times and decreases steadily
until about $1.2$ ns where the accuracy drops considerably and is
on par with the one computed using data at $M_{1}$. However, most
interestingly, there is a window of forecasting where it is possible
to obtain slightly higher accuracy with the nonlocal data at $M_{2}$
rather than using the local data at $C$. This illustrates the importance
of analyzing the transfer entropy pattern in Fig.\nobreakdash-\ref{fig:Transfer-Entropy}a,
which can allow to improve the prediction accuracy by evidencing the
spatiotemporal location of {\color{black} potential} precursors.
The forecasting accuracy at $M_{2}$ drops at $1\tau_w\simeq 1$ ns, which is more than twice the time horizon inferred
previously and also about 7.5 times larger than the Lyapunov time.
This corresponds also to the time at which the logistic regression
alone gives comparable results with the RC approach. It also relates
to the drop observed for the effective transfer entropy computed in
Fig.\nobreakdash-\ref{fig:Transfer-Entropy}b for different warning
times. This means that no useful further information can be extracted
from the input time series passed this timescale. For smaller warning
times, the reservoir is able to improve slightly the forecasting accuracy
with respect to a simpler logistic regression approach. By contrast when $1\tau_w\gg 1$ ns, very little information
 can be extracted for the prediction as testified by the low transfer
entropy computed.

In Fig.\nobreakdash-\ref{fig:Training-set-PDF-1}b,c we analyze how our
model-free approach classifies EEs depending on their actual heights.
As shown on the testing dataset histogram of heights (Fig.\nobreakdash-\ref{fig:Training-set-PDF-1}b),
despite the fact that the training sets have been balanced,
large EEs are still far less frequent than {\color{black} smaller}
ones and will therefore participate less to the training. The probability
$P(\mathrm{Event=EE}|H)$ of forecasting as an EE an event of actual
height $H$ at C, given the knowledge of a history of non-local data
at $M_{2}$ characterized with $\tau_{w}=1$ ns and $\tau_{h}=1.75$
ns is shown on Fig.\nobreakdash-\ref{fig:Training-set-PDF-1}c). In
the perfect case, the probability would evaluate
to one above $2H_{s}$ and zero below. It increases with $H$ and
generally reaches one for the largest EE heights values, while the
complementary probability $P(\mathrm{Event=NE}|H)$ goes to zero (green
and yellow histograms, respectively). This trend is true for both
the results shown in Fig.\nobreakdash-\ref{fig:Training-set-PDF-1}b,c,
obtained for two different training sets, and shows that while large
EEs are less frequent in the training dataset, their prediction accuracy
increases with their height, resulting in a usually very good prediction
for the largest EEs. We note, however, that some statistical fluctuations
can remain in the forecasting results as can be seen on the far right
of Fig.\nobreakdash-\ref{fig:Training-set-PDF-1}c) where an isolated
event has been misclassified in one realization of the train/test
datasets partitions.

\paragraph*{}
In conclusion, we have shown that a model-free approach based on reservoir
computing can successfully classify with a reasonable accuracy the
occurrence of EE in a dataset of an experimental system displaying
high dimensional spatiotemporal chaos, from the partial knowledge
of the history of the spatiotemporal field. Using the transfer entropy
concept, we identify specific spatiotemporal regions
with high information {\color{black}flow pointing to potential precursors,
which in our specific case are hidden in the dynamical fluctuations
or detection noise.} We find that while the prediction using the
local information gives generally the best accuracy, forecasting from
the nonlocal precursor {\color{black}region} can yield comparable to slightly higher accuracy
in a window of large warning times. The forecasting ability extends
to at least twice the time horizon computed from the nonlinear local
Lyapunov exponent of the system and about 7.5 times the Lyapunov time,
before dropping to a random prediction. We believe these results 
pave the way to extreme forecasting in other areas
of science, with applications to many natural systems, {\color{black}
such as in geoscience for the detection
of earthquakes where the precurors are unknwon and the spatial detection
is incomplete.}


\begin{acknowledgments}
MGC thanks for the financial support of ANID- Millenium Science Initiative
Program--ICN17\_012 (MIRO) and FONDECYT Project No. 1210353. This work was partially supported by the French Renatech network.
\end{acknowledgments}


%

\end{document}